\documentclass{ws-p8-50x6-00}

\input epsf.sty

\newcommand{\Bx}{x_{\rm B}}

\newcommand{\insertfig}[2]{\mbox{\epsfxsize=#1cm \epsfbox{#2.eps}}}

\begin{document}

\centerline{\large \bf Spin effects in deeply virtual Compton scattering}

\vspace{15mm}

\centerline{\bf A.V. Belitsky$^a$, A. Kirchner$^b$, D. M\"uller$^{b,c}$}

\vspace{10mm}

\centerline{\it $^a$C.N.\ Yang Institute for Theoretical Physics}
\centerline{\it State University of New York at Stony Brook}
\centerline{\it NY 11794-3840, Stony Brook, USA}

\vspace{5mm}

\centerline{\it $^b$Institut f\"ur Theoretische Physik,
                Universit\"at Regensburg}
\centerline{\it D-93040 Regensburg, Germany}

\vspace{5mm}

\centerline{\it $^c$Fachbereich Physik, Universit\"at Wuppertal,}
\centerline{\it D-42097 Wuppertal, Germany}

\vspace{15mm}

\centerline{\bf Abstract}

\vspace{0.5cm}

\noindent
We consider the azimuthal angle dependence in the cross section
of the hard leptoproduction of a photon on a nucleon target. We show that
this dependence allows to define observables that isolate the twist-two and
twist-three sectors in the deeply virtual Compton scattering amplitude. All
twist-two and twist-three Compton form factors can be extracted from measurements
of the charge odd part of the polarized cross section and give access to all
generalized parton distributions.

\vspace{2cm}

\centerline{\it Talk given at the}
\centerline{\it IX International Workshop on Deep Inelastic Scattering}
\centerline{\it  Bologna, 27 April - 1 May 2001 }


\newpage

\title{Spin effects in deeply virtual Compton scattering}

\author{A.V. Belitsky}

\address{C.N.\ Yang Institute for Theoretical Physics\\
         State University of New York at Stony Brook\\
         NY 11794-3840, Stony Brook, USA}

\author{A. Kirchner}

\address{Institut f\"ur Theoretische Physik, Universit\"at Regensburg\\
         D-93040 Regensburg, Germany }

\author{D. M\"uller}

\address{Fachbereich Physik, Universit\"at Wuppertal\\
         D-42097 Wuppertal, Germany}

\maketitle

\thispagestyle{empty}

\abstracts{We consider the azimuthal angle dependence in the cross section
of the hard leptoproduction of a photon on a nucleon target. We show that
this dependence allows to define observables that isolate the twist-two and
twist-three sectors in the deeply virtual Compton scattering amplitude. All
twist-two and twist-three Compton form factors can be extracted from measurements
of the charge odd part of the polarized cross section and give access to all
generalized parton distributions.}

\section{Introduction}

The hard leptoproduction of a photon  is a promising process to access a new
type of non-perturbative distribution functions, the so-called generalized
parton distributions (GPDs) \cite{MueRobGeyDitHor94,Ji96a,Rad96}. In this
process the partonic content of the hadron is studied by an electromagnetic
probe and is free from the fragmentation in the final state present e.g.\ in
the hard leptoproduction of a meson. Although the leptoproduction of a photon
contains the contaminating Bethe-Heitler (BH) process in addition to the
deeply virtual Compton scattering (DVCS), their interference gives a particular
term since it provides a large number of observables that linearly depend
on GPDs \cite{Ji96a,DieGouPirRal97,BelMueNieSch00}.

The GPDs are defined as non-diagonal matrix elements of light-ray operators
sandwiched between initial and final hadronic states with different momenta
and possibly different quantum numbers:
\begin{eqnarray}
\label{Def-GPDs-gen}
q(x, \eta, \Delta^2 | Q^2)
=
\int \frac{d \kappa}{2 \pi}
{\rm e}^{i x \kappa (n\cdot P)}
\langle P_2 |
\bar \psi (- \kappa n)
{\mit\Gamma}
[- \kappa, \kappa]
\psi (\kappa n)
| P_1 \rangle_{|\mu^2=Q^2} .
\end{eqnarray}
Here $n_\mu$ is a light-like vector and ${\mit\Gamma}$ denotes the Dirac and
flavour structure. They depend on the (partonic) momentum fraction $x$ in the
s-channel, the longitudinal momentum fraction $\eta$ in the $t$-channel,
called skewedness parameter, the momentum transfer $\Delta^2$, the resolution
scale $Q^2$, and the helicities of the hadrons. Their evolution arises from
the renormalization group equations of the light-ray operators and is known to
next-to-leading order (NLO) of perturbation theory.

As we see the GPDs have a complex structure. On the other side their
definition implies also that they encode new information about the
non-perturbative QCD sector. This information can not be obtained from other
measurable non-perturbative quantities, i.e.\ parton densities or hadronic
distribution amplitudes. The second moment of certain GPDs is directly
related to gravitational form factors and could give some insight in the
spin structure of the nucleon\cite{Ji96}. It is also very interesting, that
their first moments give us partonic form factors. Moreover, the GPDs provide
us with a link between different exclusive and inclusive quantities. In the
forward case the non-helicity flip GPDs reduce to the parton densities. In
the `exclusive' region of the phase space, i.e.\ $|x| \le |\eta|$, they are
related to distribution amplitudes of mesonic like states, while in the
`inclusive' region they are analogous to parton densities. They are also
related to hadronic wave functions by a Drell-Yan-West overlap type
representation\cite{DieFelJakKro00,BroDieHwa00}.

\section{The general azimuthal angular dependence}

\begin{figure}[h]
\unitlength1mm
\begin{center}
\mbox{
\begin{picture}(0,28)(50,0)
\put(0,-4){\insertfig{9}{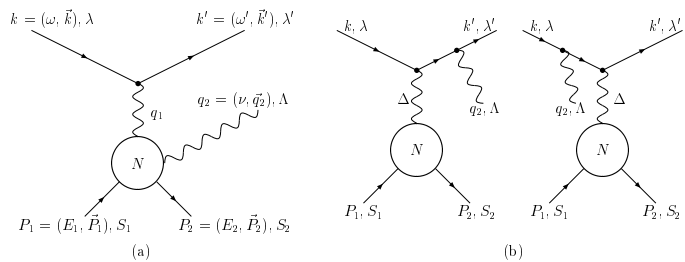}}
\end{picture}
}
\end{center}
\caption{\label{electroprod} The virtual Compton scattering amplitude (a)
and the Bethe-Heitler (b) process.}
\end{figure}

As mentioned before the hard leptoproduction of a photon contains two
processes, the Bethe-Heitler and the DVCS (see Fig. \ref{electroprod}).
\vspace{-0.5cm}
The amplitude of the former one is parametrized for a nucleon target by the
Pauli- and Dirac-form factors $F_1(\Delta^2)$ and $F_2(\Delta^2)$, which appear
in the hadronic current
\begin{eqnarray}
\label{def-HadCur}
J_\nu = \bar U (P_2,S_2)
\left\{ F_1(\Delta^2) \gamma_\nu
+ i F_2(\Delta^2) \sigma_{\nu\tau} \frac{\Delta^\tau}{2M}
\right\} U (P_1,S_1),
\end{eqnarray}
where $ \Delta = P_2 - P_1 = q_1 - q_2$.
 The amplitude of the DVCS process is given in terms of the hadronic tensor
that reads at twist-two level \cite{BelMueNieSch00}
\begin{eqnarray}
\label{HadronicTensor}
T_{\mu\nu} (q, P, \Delta)
=
- {\cal P}_{\mu\sigma} g_{\sigma\tau} {\cal P}_{\tau\nu}
\frac{q \cdot V_1}{P \cdot q}
- {\cal P}_{\mu\sigma} i\epsilon_{\sigma \tau q \rho} {\cal P}_{\tau\nu}
\frac{A_{1\, \rho}}{P \cdot q} + \cdots,
\end{eqnarray}
where ${\cal P}_{\mu\nu}$ ensures gauge invariance, $q = (q_1 + q_2)/2$ and
$P = P_1 +P_2.$ Analogous to Eq.\ (\ref{def-HadCur}), we decompose below the
vector $V_1$ and axial-vector $A_1$ amplitudes into form factors, the so-called
Compton form factors (CFFs).

Let us start with the azimuthal angle dependence of the interference term
for an unpolarized lepton scattering on a polarized nucleon in the target
rest frame where $q_1 = (q_1^0,0,0,-|q_1^3|)$. The interference term can be
written as a contraction with a leptonic tensor $L_{\mu\nu\al}$:
\begin{eqnarray}
{\cal I} (S)
=  \pm  \frac{e^6}{\Delta^2 {\cal Q}^2}
L_{\mu\nu\al}
 \sum_{S_2} T^{\mu\nu} {J^\alpha}^\ast
+ {\rm h.c.}
\quad\mbox{for\ }
\left\{ {+ \mbox{\ for\ } e^- \atop - \mbox{\ for\ } e^+ } \right. .
\end{eqnarray}
Here $L_{\mu\nu\al}$ is built from the leptonic four-vectors $k$ and $k'$.
The longitudinal polarization vector $S_L$, having only a $z$-component, can
be constructed from $q_1$ and $P_1$. In this case the azimuthal angular
dependence $\phi = \phi_N - \phi_l$, where $\phi_N$ ($\phi_l$) is the azimuthal
angle of the outgoing nucleon (lepton), can only arise from a contraction of
$k$ with $\Delta$ (or $P$) and from $\epsilon_{kqP\Delta}$. Therefore, the
highest moment in the Fourier series is $\cos/\sin 3\phi$:
\begin{eqnarray}
{\cal I} =
\pm \frac{e^6}{\Bx y^3 {\cal P}_1 (\phi) {\cal P}_2(\phi) \Delta^2}
\left\{ c_0^{\cal I}+  \sum_{n = 1}^3
\left[
c_n^{\cal I} \cos(n\phi) +  s_n^{\cal I} \sin(n\phi)
\right]\right\} .
\end{eqnarray}
The additional $\phi$ dependence in the prefactor is induced by the (scaled)
lepton propagators ${\cal P}_1 \equiv (k - q_2)^2/{\cal Q}^2 $ and ${\cal
P}_2 \equiv (k - \Delta)^2/{\cal Q}^2$ of the BH process, where ${\cal
Q}^2\equiv-q_1^2$. The dimensionless Fourier coefficients depend on the
beam and target polarization $\lambda$ and $\Lambda$, respectively. A more
detailed analysis shows that $c_1^{\cal I}$ ($s_1^{\cal I}$) and
$c_3^{\cal I}$ ($s_3^{\cal I}$) arise from the leading twist-two sector,
while $c_0^{\cal I}$ and $c_2^{\cal I}$ ($s_2^{\cal I}$) appear at the
twist-three level and are additionally suppressed by $1/{\cal Q}$. The
coefficient $c_3^{\cal I}$ ($s_3^{\cal I}$) is caused by the spin-flip of
the photon by two units and is, thus, related to the tensor gluon
operators. Such a contribution is perturbatively suppressed by the strong
coupling $\alpha_s$. In the case of a transversely polarized target,
$S_T = (0,\cos\Phi,\sin\Phi,0)$, the coefficients can be additionally
decomposed with respect to their $\varphi = \phi_N - \Phi$ dependence:
\begin{eqnarray}
c_n^{\cal I}(\varphi)
=
\delta_{n0} c_{\rm unp}^{\cal I}
+
\cos\varphi\; c_{n+}^{\cal I}
+
\sin\varphi\; s_{n-}^{\cal I} , \
s_n^{\cal I}(\varphi)
=
\cos\varphi\; s_{n+}^{\cal I}
+
\sin\varphi\; c_{n-}^{\cal I} .
\end{eqnarray}
All the coefficients $c_n^{\cal I}$ ($s_n^{\cal I}$) appear in double-spin
(single-spin) flip experiments and are given by the real (imaginary) part of
CFFs. To extract them from the charge odd part of the cross section one has
to form appropriate $\cos/\sin (m \phi)$ moments with the overall weight
${\cal P}_1(\phi){\cal P}_2(\phi)$.

The charge even sector is given by the sum of the squared DVCS and BH term.
In certain kinematical regions it is possible to access the former by
subtracting the latter. The definition of the squared DVCS term
\begin{eqnarray}
|{\cal T}_{\rm DVCS}|^2
= - \frac{e^6}{{\cal Q}^4} L_{\mu\nu}  \sum_{S_2}
{T^\mu_{\;\;\,\al}} {T^{\mu\al}}^\ast
\end{eqnarray}
allows an analysis analogous to the one given above:
\begin{eqnarray}
|{\cal T}_{\rm DVCS}|^2
&=& \frac{e^6}{y^2 {\cal Q}^2}
\sum_{n=0}^2
\left[
c^{\rm DVCS}_n \cos (n\phi) + s^{\rm DVCS}_n \sin (n \phi)
\right] .
\end{eqnarray}
Here the coefficients $c^{\rm DVCS}_0$, $c^{\rm DVCS}_2$, and $s^{\rm
DVCS}_2$ ($c^{\rm DVCS}_1$ and $s^{\rm DVCS}_1$) appear at twist-two
(twist-three) level. However, $c^{\rm DVCS}_2$ and $s^{\rm DVCS}_2$
arise from the interference of the tree level contribution with the
gluonic transversity and, thus, they are perturbatively suppressed
by $\alpha_s$.

\section{Predictions for twist-two and -three observables}

We sketch now the calculation of the Fourier coefficients. First we
parametrize the vector and axial-vector contributions in four CCFs ${\cal F}
= \{ {\cal H},{\cal E},\widetilde{{\cal H}},\widetilde{{\cal E}}\}$:
\begin{eqnarray}
\label{Def-V1}
V_{1\, \rho} =
P_\rho  \frac{q\cdot h}{q\cdot P} {\cal H}
+
P_\rho  \frac{q\cdot e}{q\cdot P} {\cal E}
+ \cdots , \
A_{1\, \rho} =
P_\rho  \frac{q\cdot \widetilde h}{q\cdot P} \widetilde{\cal H}
+
P_\rho  \frac{q\cdot \widetilde e}{q\cdot P} \widetilde{\cal E}
+ \cdots ,
\end{eqnarray}
where $\{h, e,\widetilde h,\widetilde e\}$ are vector, tensor, axial-vector,
and pseudo-scalar Dirac bilinears and the ellipsis stand for higher twist
contributions. The CCFs are given as convolutions of a hard-scattering
amplitude, calculable in perturbation theory, with the GPDs and read in
leading order (LO)
\begin{eqnarray}
\label{Pre-LO-DVCS}
&&\!\!\!\!\!\!\!\!\!\!\!\!\!
{\cal F}(\xi, \Delta^2 | {\cal Q}^2)
= \!\!\sum_{i = u, d, s} \int_{-1}^{1}\!\! d x
\left( \frac{Q_i^2}{\xi - x - i 0} \pm \{ x \to -x\}
\right)
q_i (x, - \xi, \Delta^2 | {\cal Q}^2),
\end{eqnarray}
where the scaling variable $\xi\approx \Bx/(2-\Bx)$ is related to the
Bjorken variable $\Bx = {\cal Q}^2/q_1\cdot P_1$ and $Q_i$ are the charge
fractions of quarks participating in the scattering. The complete NLO
corrections to the twist-two sector are worked out \cite{BelMueSchNie99},
while the twist-three sector is completely known to LO accuracy
\cite{BelMue00a,BelKirMueSch01}.

The Fourier coefficients for unpolarized (unp), longitudinally (LP), and
transversely polarized (TP) target are expressed in terms of combinations
\begin{eqnarray}
&&\!\!\!{\cal C}^{\cal I}_{\rm unp}
=
F_1 {\cal H} + \frac{\Bx}{2 - \Bx}
(F_1 + F_2) \widetilde {\cal H}
-
\frac{\Delta^2}{4M^2} F_2 {\cal E},
\\
&&\!\!\!{\cal C}^{\cal I}_{\rm LP}
=
\frac{\Bx }{2 - \Bx}(F_1 + F_2) \left({\cal H} +
\frac{\Bx}{2} {\cal E} \right)
+ F_1 \widetilde{\cal H}
- \frac{\Bx}{2 - \Bx}
\left(
\frac{\Bx}{2} F_1 + \frac{\Delta^2}{4 M^2} F_2 \right)
\widetilde{\cal E},
\nonumber\\
&&\!\!\!{\cal C}^{\cal I}_{{\rm TP}+}
=
(F_1 + F_2) \left\{ \frac{\Bx^2}{2 - \Bx} \left( {\cal H} +
\frac{\Bx}{2} {\cal E} \right) +  \frac{\Bx \Delta^2}{4 M^2}{\cal E} \right\}
\nonumber\\
&&
-
F_1
\frac{\Bx^2}{2 - \Bx} \left( \widetilde{\cal H} +
\frac{\Bx}{2} \widetilde{\cal E}\right)
+ \frac{1 - \Bx}{2 - \Bx} \frac{\Delta^2}{M^2} F_2
\widetilde{\cal H}
- \frac{\Delta^2}{4M^2} \left( \Bx F_1 + \frac{\Bx^2}{2 - \Bx} F_2 \right)
\widetilde{\cal E}
\nonumber\\
&&\!\!\!{\cal C}^{\cal I}_{{\rm TP}-}
=
\frac{1}{2 - \Bx} \left(\Bx^2 F_1 - (1 - \Bx) \frac{\Delta^2}{M^2} F_2  \right) {\cal H}
+
\Bigg\{ \frac{\Bx^2}{2 - \Bx} F_1
\nonumber\\
&&+ \frac{\Delta^2}{4 M^2}
\left( (2 - \Bx) F_1 + \frac{\Bx^2}{2 - \Bx} F_2 \right) \Bigg\} {\cal E}
-\frac{\Bx^2}{2 - \Bx} (F_1 + F_2)  \left( \widetilde{\cal H} +
\frac{\Delta^2}{4 M^2}\widetilde{\cal E} \right) .
\nonumber
\end{eqnarray}
Note that these relations can be inverted to determine the set ${\cal F}$
from the set ${\cal C}= \{{\cal C}_{\rm unp}, {\cal C}_{\rm LP}, {\cal
C}_{\rm TP+}, {\cal C}_{\rm TP-} \}$. For instance, in the case of an
unpolarized target the $\cos/\sin\phi$ (twist-two) and $\cos/\sin 2 \phi$
(twist-three) coefficients read
\begin{eqnarray}
\label{Res-IntTer-2}
\left\{{c^{\cal I}_{1, \rm unp} \atop s^{\cal I}_{1, \rm unp}}\right\}
&=&
8 K
\left\{  {-(2 - 2y + y^2) \atop \lambda y (2-y)} \right\}
\left\{{{\rm Re} \atop {\rm Im} } \right\}
{\cal C}^{\cal I}_{\rm unp}\left({\cal F} \right),
\\
\label{Res-IntTer-3}
\left\{ c^{\cal I}_{2, \rm unp} \atop  s^{\cal I}_{2, \rm unp} \right\}
&=&
\frac{16 K^2}{2 - \Bx}  \left\{ { -(2 - y) \atop \lambda y} \right\}
\left\{{{\rm Re} \atop {\rm Im} } \right\}
{\cal C}^{\cal I}_{\rm unp}
\left(
{\cal F}^{\rm eff}\right) ,
\end{eqnarray}
where $K \propto \left\{ - \Delta^2/{\cal Q}^2 \left( 1 -
\Delta^2_{\rm min}/\Delta^2 \right) \right\}^{1/2}$ with
$-\Delta^2_{\rm min}$ being the minimal value of $-\Delta^2$. In the
twist-three sector a new set ${\cal F}^{\rm eff}$ of four twist-three
CCFs arise. They are partly given by the twist-two GPDs through the
Wandzura-Wilczek relation but  are also sensitive to new dynamical effects
arising from antiquark-gluon-quark correlations. The four twist-three
coefficients $c^{\cal I}_0$ are of purely kinematical origin and are given
in terms of twist-two CFFs. Up to different kinematical prefactors, relations
(\ref{Res-IntTer-2},\ref{Res-IntTer-3}) hold true for polarized targets too.

The squared DVCS term can be calculated in the same manner. The functional
dependence at twist-two and -three levels should be universal, too.

\section{Discussions and conclusions}

We defined appropriate observables in the hard leptoproduction of a photon
that separate the twist-two and -three sectors. We expect that higher twist
contributions will only affect them by $1/{\cal Q}^2$ suppressed
corrections. Facilities that have lepton beams of both charges can
access the interference term. In combination with single ($\sin n\phi$) and
double spin flip experiments ($\cos n\phi$) and by forming appropriate
moments with respect to azimuthal angles one can extract all possible
twist-two and -three CFFs:
\begin{itemize}
\item unpolarized beam and unpolarized target
\hspace{1.4cm} $\to {\rm Re} {\cal C}^{\cal I}_{\rm unp}$
\vspace{-0.3cm}
\item double spin flip with longitudinal polarized target
\hspace{0.1cm} $\to {\rm Re} {\cal C}^{\cal I}_{\rm LP}$
\vspace{-0.3cm}
\item double spin flip with transverse polarized target
\hspace{0.4cm} $\to {\rm Re} {\cal C}^{\cal I}_{\rm TP \pm}$
\vspace{-0.3cm}
\item polarized beam and unpolarized target
\hspace{1.8cm} $\to {\rm Im} {\cal C}^{\cal I}_{\rm unp}$
\vspace{-0.3cm}
\item unpolarized beam and longitudinal polarized target
$\to {\rm Im} {\cal C}^{\cal I}_{\rm LP}$
\vspace{-0.3cm}
\item unpolarized beam and transverse polarized target
\hspace{0.3cm}$\to {\rm Im} {\cal C}^{\cal I}_{\rm TP \pm}$
\end{itemize}
Then the coefficients $c_0^{\cal I}$ and $c^{\rm DVCS}_0$ of the
interference and the squared DVCS term can serve for an experimental
consistency check at twist-two level while $c^{\rm DVCS}_1$ and $s^{\rm
DVCS}_1$ allow the same at twist-three level.

The second step would be to extract information about the GPDs from such
measurements by means of Eq.\ (\ref{Pre-LO-DVCS}). In LO approximation one
can directly extract the shape of the GPDs at the diagonal $x = \pm \xi$
from the imaginary part of the CCFs. However, beyond the tree level this
feature is lost. Indeed, the (magnitude of the) theoretical predictions is
sensitive to special features, but, in general not to the shape of GPDs, so
that experimental measurements can give constraints for models of GPDs. A
crucial issue for this is the contamination of Eq.\ (\ref{Pre-LO-DVCS}) by
perturbative and power suppressed contributions. The former ones have been
evaluated at NLO in the twist-two sector and model dependent numerical
estimates show that they can be large. The study of the latter has been
just started with a consideration of the target mass corrections.

\end{document}